ArticleArticle

Statistical Methods in Medical Research
0(0) 1–21
© The Author(s) 2016
Reprints and permissions:
sagepub.co.uk/journalsPermissions.nav
DOI: 10.1177/0962280216669183
smm.sagepub.com
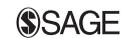

# Optimally estimating the sample mean from the sample size, median, mid-range, and/or mid-quartile range

Dehui Luo,[1] Xiang Wan,[2] Jiming Liu[2] and Tiejun Tong[1]## Abstract

The era of big data is coming, and evidence-based medicine is attracting increasing attention to improve decision making in medical practice via integrating evidence from well designed and conducted clinical research. Meta-analysis is a statistical technique widely used in evidence-based medicine for analytically combining the findings from independent clinical trials to provide an overall estimation of a treatment effectiveness. The sample mean and standard deviation are two commonly used statistics in meta-analysis but some trials use the median, the minimum and maximum values, or sometimes the first and third quartiles to report the results. Thus, to pool results in a consistent format, researchers need to transform those information back to the sample mean and standard deviation. In this article, we investigate the optimal estimation of the sample mean for meta-analysis from both theoretical and empirical perspectives. A major drawback in the literature is that the sample size, needless to say its importance, is either ignored or used in a stepwise but somewhat arbitrary manner, e.g. the famous method proposed by Hozo et al. We solve this issue by incorporating the sample size in a smoothly changing weight in the estimators to reach the optimal estimation. Our proposed estimators not only improve the existing ones significantly but also share the same virtue of the simplicity. The real data application indicates that our proposed estimators are capable to serve as "rules of thumb" and will be widely applied in evidence-based medicine.

## Keywords
Median, meta-analysis, mid-range, mid-quartile range, optimal weight, sample mean, sample size
## 1 Introduction

The concept of evidence-based medicine was introduced in 1992 by Guyatt et al.,[1] which was intended to improve decision making in medical practice and has risen to be regarded as the gold standard for healthcare and medicine. It integrates findings from several independent studies of the same clinical question and then use statistical techniques to combine the results together so that proper decisions to cure patients could be made eventually. Meta-analysis plays a crucial role in evidence-based medicine to help researchers summarize data comprehensively.[2] To statistically combine data from multiple studies, the first step is to determine a summary measure as well as its corresponding statistics. In most of the studies, the sample mean and standard deviation are two commonly used statistics in the data analysis. However, instead of directly reporting the sample mean and standard deviation, the median, the first and third quartiles, the minimum and maximum values are often recorded in clinical trial studies. As a result, when proceeding meta-analysis, people need to transform these information to the sample mean and standard deviation in order to conduct further analysis.

To transform the data, Hozo et al.[3] used inequalities to establish some estimators for the sample mean and variance. They were the first to suggest methodology for this estimation problem. Their proposed method is simple

[1]Department of Mathematics, Hong Kong Baptist University, Hong Kong
[2]Department of Computer Science, Hong Kong Baptist University, Hong Kong

**Corresponding authors:**
Xiang Wan, Department of Computer Science, Hong Kong Baptist University, Hong Kong.
Email: xwan@comp.hkbu.edu.hk
Tiejun Tong, Department of Mathematics, Hong Kong Baptist University, Hong Kong.
Email: tongt@hkbu.edu.hk
Downloaded from smm.sagepub.com at CORNELL UNIV on September 30, 2016

and has been widely adopted in the scenario where only the sample median, extremum, and sample size are reported. Recently, Wan et al.[4] pointed out that Hozo et al.'s method has some serious drawbacks and, in particular, is less accurate for the estimation of the sample variance. In view of this, they introduced a quartile method to improve the sample standard deviation estimation. They have further extended the new methodology to two other frequently encountered scenarios in reporting clinical trial results. Through simulation studies, they have demonstrated that their newly proposed methods greatly outperform the existing methods including Hozo et al.[3] and Bland.[5]

Wan et al.[4] had fully discussed and consummated the approaches in estimating the sample standard deviation under different conditions. For the estimation of the sample mean, they simply followed the estimation methods in Hozo et al.[3] and Bland.[5] These existing methods, however, suffer from some major limitations due to the insufficient use of the information in the sample size. In this article, we propose some new methods by incorporating the sample size in a smoothly changing weight to reach the optimal estimation of the sample mean. The proposed methods not only improve the existing estimators significantly but also share the same virtue of the simplicity. We believe our proposed estimators will serve as "rules of thumb" for the sample mean estimation in meta-analysis.

## 2 Sample mean estimation

For the sake of consistency, we follow essentially the same notations as those in Hozo et al.[3] and Wan et al.[4] Specifically, we let $n$ be the sample size and denote the five-number summary for the data as

$$a = \text{the minimum value},$$
$$q_1 = \text{the first quartile},$$
$$m = \text{the median},$$
$$q_3 = \text{the third quartile},$$
$$b = \text{the maximum value}.$$

In clinical trial reports, the five-number summary for the data may not be provided in full. We consider the three scenarios that are most frequently encountered:

$$\mathcal{S}_1 = \{a, m, b; n\},$$
$$\mathcal{S}_2 = \{q_1, m, q_3; n\},$$
$$\mathcal{S}_3 = \{a, q_1, m, q_3, b; n\}$$

According to Triola,[6] we refer to $(b - a)$ as the range, $(a + b)/2$ as the mid-range, $(q_3 - q_1)$ as the interquartile range, and $(q_1 + q_3)/2$ as the mid-quartile range. As a common practice, the range and the interquartile range are often used to measure the standard deviation, whereas the mid-range and the mid-quartile range are used to measure the center (or mean) of the population.

### 2.1 Existing methods

#### 2.1.1 Hozo et al.'s method for $\mathcal{S}_1 = \{a,m,b; n\}$

Scenario $\mathcal{S}_1$ represents the situation where the median, the minimum, the maximum, and the sample size are given in a study. Hozo et al.[3] were the first to address the sample mean estimation problem. By inequalities, they proposed the following estimator for the sample mean:

$$\bar{X} \approx \begin{cases} (a + 2m + b)/4 & n \leq 25, \\ m & n > 25 \end{cases} \qquad (1)$$

Although very easy to implement, we note that the estimator (1) may not be sufficiently accurate as it incorporates the sample size in a stepwise manner. The sample mean estimation has a sudden change from $m$ to $(a + 2m + b)/4$ when the sample size reduces to 25. This change might lead to a less precise estimation. For example, when the sample size is changed from 26 to 25, the estimated sample mean might be a lot different than the actual one because of the "jump" in the estimator. In contrast, within the respective interval of $n > 25$ or $n \leq 25$, the sample mean estimation is independent of the sample size and the information in the sample size is completely ignored. As a consequence, such an estimate may not be reliable for practical use. This motivates us to consider an improved estimation of the sample mean by incorporating the sample size in a smoothly changing manner.

### 2.1.2 Wan et al.'s method for $S_2 = \{q_1, m, q_3; n\}$

Scenario $S_2$ reports the first and third quartiles instead of the minimum and the maximum, together with the median and the sample size. Other than the sample range, i.e. the difference between the minimum and the maximum, the interquartile range is usually less sensitive to outliers and hence is also popularly reported in clinical trial studies. For scenario $S_2$, Wan et al.[4] proposed to estimate the sample mean by

$$\bar{X} \approx \frac{q_1 + m + q_3}{3} \quad (2)$$

It is evident that the sample size information is not used in their proposed estimation. We also note that an equal weight is assigned to each summary statistic in the estimator (2). In particular, the weight for the median is 1/3 in scenario $S_2$ compared with 1/2 in scenario $S_1$. Hence, it would also be of interest to investigate if a smoothly changing manner is needed for assigning the appropriate weights to the median and the two quartiles with respect to the sample size.

### 2.1.3 Bland's method for $S_3 = \{a, q_1, m, q_3, b; n\}$

Scenario $S_3$ is a combination of the scenarios $S_1$ and $S_2$. It assumes that the five-number summary of the data are all given for further analysis. Following the same idea in Hozo et al.,[3] Bland[5] proposed the following estimator for the sample mean:

$$\bar{X} \approx \frac{a + 2q_1 + 2m + 2q_3 + b}{8} \quad (3)$$

Once again, the sample size information is not used in the estimation of the sample mean. The estimator assigns an equal weight to $q_1$, $m$, and $q_3$, respectively, and another equal weight to $a$ and $b$, respectively. Similar to the other two scenarios, we will investigate if a smoothly changing manner is needed for assigning the appropriate weights to the five-number summary of the data with respect to the sample size.

## 2.2 Improved methods

Let $X_1, X_2, \ldots, X_n$ be a random sample of size $n$ from the normal distribution $N(\mu, \sigma^2)$, and $X_{(1)} \leq X_{(2)} \leq \cdots \leq X_{(n)}$ be the ordered statistics of the sample. For simplicity, we assume that the sample size $n = 4Q + 1$ with $Q \geq 1$ being a positive integer. With the above notations, we have $a = X_{(1)}$, $q_1 = X_{(Q+1)}$, $m = X_{(2Q+1)}$, $q_3 = X_{(3Q+1)}$, and $b = X_{(n)} = X_{(4Q+1)}$. For convenience, let also $X_i = \mu + \sigma Z_i$, or equivalently, $X_{(i)} = \mu + \sigma Z_{(i)}$ for $i = 1, \ldots, n$. Then, $Z_1, Z_2, \ldots, Z_n$ follows the standard normal distribution $N(0, 1)$, and $Z_{(1)} \leq Z_{(2)} \leq \cdots \leq Z_{(n)}$ are the ordered statistics of the sample $\{Z_1, \ldots, Z_n\}$.

### 2.2.1 Improved estimation of the sample mean in $S_1 = \{a, m, b; n\}$

Following the discussion in "Hozo et al.'s method for $S_1 = \{a, m, b; n\}$" section, we propose to estimate the sample mean by

$$\bar{X}(w) = w\left(\frac{a+b}{2}\right) + (1-w)m \quad (4)$$

where $w$ is the weight assigned to the mid-range $(a+b)/2$, and the remaining weight $1 - w$ is assigned to the median $m$. The proposed estimator (4) is a weighted average of the mid-range and the median, where both quantities are the measures of center for the population. In the special case if we take $w = 1/2$ for $n \leq 25$ and $w = 0$ for $n > 25$, the proposed estimator reduces to the estimator (1) proposed by Hozo et al.[3] Such an allocation of the weight is somewhat arbitrary and can be less reliable.

We consider to solve the issue by incorporating the sample size in a smoothly changing manner. That is, we consider the weight $w = w(n)$ as a function of the sample size. Then from the decision-making point of view, we define the optimal weight $w_{\text{opt}} = w_{\text{opt}}(n)$ to be the weight that minimizes the expected loss function of the estimator. In this article, we consider the squared loss function $L(\bar{X}(w), \mu) = (\bar{X}(w) - \mu)^2$, then accordingly, the expected loss function is the commonly used mean squared error (MSE) of the estimator. By Theorem 1 in Appendix 2, the proposed estimator (4) is an unbiased estimator of the true mean $\mu$. Hence, we have $\text{MSE}(\bar{X}(w)) = (w^2/4)\text{Var}(a+b) + (1-w)^2\text{Var}(m) + w(1-w)\text{Cov}(a+b,m)$. Note that the MSE function is a quadratic function of $w$ and has a unique minimum value on $[0, 1]$.

To derive the optimal weight, we take the first derivative of MSE with respect to $w$ and set the result equal to zero. It gives the optimal weight as

$$w_{\text{opt}}(n) = \frac{4\text{Var}(m) - 2\text{Cov}(a+b,m)}{\text{Var}(a+b) + 4\text{Var}(m) - 4\text{Cov}(a+b,m)}$$

Recall that $a = \mu + \sigma Z_{(1)}$, $b = \mu + \sigma Z_{(n)}$ and $m = \mu + \sigma Z_{(2Q+1)}$. Together with the symmetry of the standard normal distribution, we can represent the optimal weight as

$$w_{\text{opt}}(n) = \frac{K(n)}{K(n)+1} \qquad (5)$$

where $K(n) = 2[E(Z_{(2Q+1)}^2) - E(Z_{(1)}Z_{(2Q+1)})]/[E(Z_{(1)}^2) + E(Z_{(1)}Z_{(n)}) - 2E(Z_{(1)}Z_{(2Q+1)})]$. The derivation of (5) is given in the proof of Theorem 1 in Appendix 2. It is clear that the optimal weight $w_{\text{opt}}(n)$ is independent of $\mu$ and $\sigma^2$ and is only a function of $n$.

To explore the behavior of the optimal weight, we use the statistical software $R$ to numerically compute the values of $w_{\text{opt}}(n)$ for $n$ from 5 to 101 and plot them in the top panel of Figure 1. We observe that $w_{\text{opt}}(n)$ decreases as $n$ increases, in particular, the optimal weight reduces to about 0.1 when $n = 101$. When the sample size is large or very large, the estimator will assign most of the weight to the median as it provides a more robust estimate for the measure of center compared with the mid-range. In fact, as mentioned in Triola,[6] the mid-range is rarely used in practice as, from an asymptotic point of view, it lacks efficiency and robustness as an estimator. When the sample size is small, however, a well-designed combination of the mid-range and the median may provide a better estimation of the sample mean compared with only using the median. In addition, we note that the optimal weight $w_{\text{opt}}(n)$ is about 0.25 when $n = 25$. This explains why in a stepwise manner with $w = 0$ and $w = 0.5$ being the only two options, Hozo et al.[3] suggested to take $w = 0.5$ when $n \leq 25$ and $w = 0$ when $n > 25$.

Note that the optimal weight $w_{\text{opt}}(n)$ in (5) may not be readily accessible for practitioners as it involves some complicated statistical computation. In what follows, we propose an approximation formula for $w_{\text{opt}}(n)$ and then display the final estimator of the sample mean as an "rule of thumb" for practical use. To approximate $K(n)$, we consider the simple power function $K(n) = c_1 n^{c_2}$. Using the observed true weights in the top panel of Figure 1, we figure out that the best coefficients are about $c_1 = 4$ and $c_2 = -0.75$. This leads to the approximated optimal weight as

$$\tilde{w}_{\text{opt}}(n) \approx \frac{4}{4 + n^{0.75}} \qquad (6)$$

For comparison, we also display the approximated optimal weights (6) and the weights proposed by Hozo et al. in Figure 1. It is evident that the approximated optimal weights provide a nearly perfect match to the true optimal weights, in particular for the sample size ranging from 5 to 101.

Finally, by plugging the approximation formula (6) into the estimator (4), we propose the estimator for Scenario $\mathcal{S}_1$ as

$$\bar{X}(w) \approx \left(\frac{4}{4+n^{0.75}}\right)\frac{a+b}{2} + \left(\frac{n^{0.75}}{4+n^{0.75}}\right)m \qquad (7)$$

The performance of (7) is evaluated in Section 3, together with its numerical comparison with the estimation method in Hozo et al.[3]

### 2.2.2 Improved estimation of the sample mean in $\mathcal{S}_2 = \{q_1, m, q_3; n\}$

For scenario $\mathcal{S}_2$, we propose the new estimator for the sample mean as

$$\bar{X}(w) = w\left(\frac{q_1+q_3}{2}\right) + (1-w)m \qquad (8)$$

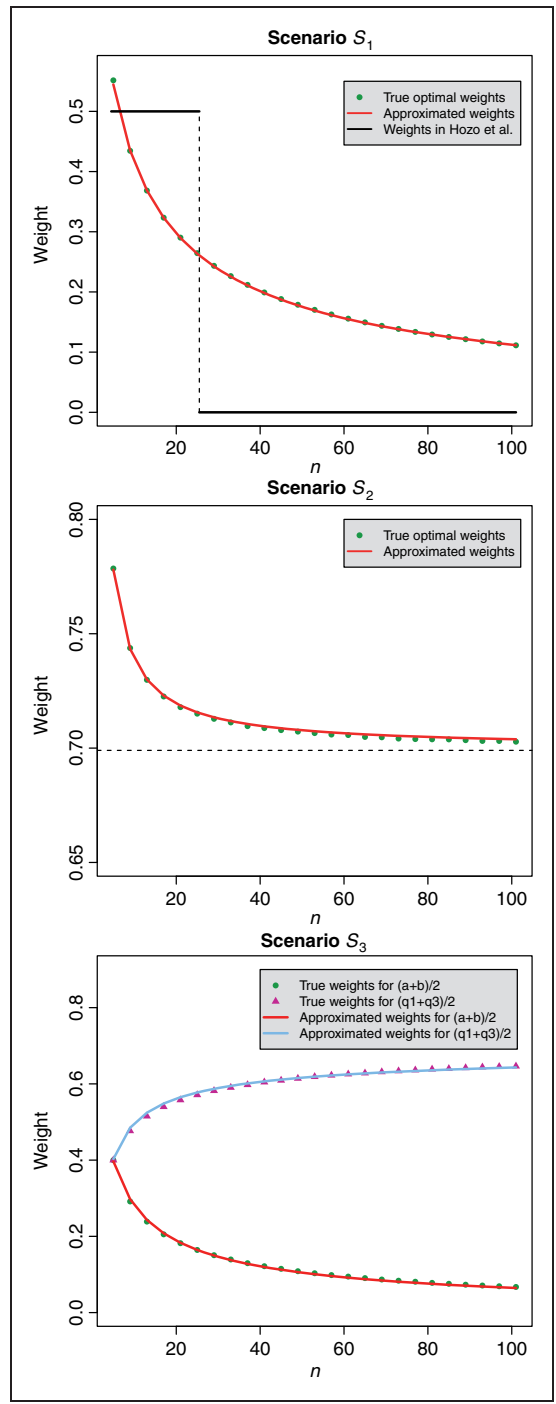

**Figure 1.** The true optimal weights (simulated using the statistical software R), and the approximated optimal weights for scenario $\mathcal{S}_1$, $\mathcal{S}_2$, and $\mathcal{S}_3$, respectively.

where $w$ and $1 - w$ are the weights assigned to the mid-quartile range $(q_1 + q_3)/2$ and the median $m$. The new estimator is a weighted average of the mid-quartile range and the median. It is worth mentioning that the mid-quartile range is also a measure of the population center, which is the numerical value midway between the first and third quartiles.[7] In addition, by taking $w = 2/3$ for all $n$, the new estimator reduces to the estimator (2) proposed by Wan et al.[4]

Following the similar arguments as in "Improved estimation of the sample mean in $\mathcal{S}_1 = \{a, m, b; n\}$" section, the optimal weight for the mid-quartile range that minimizes the MSE function is obtained as

$$w_{\text{opt}}(n) = \frac{4\text{Var}(m) - 2\text{Cov}(q_1 + q_3, m)}{\text{Var}(q_1 + q_3) + 4\text{Var}(m) - 4\text{Cov}(q_1 + q_3, m)} \qquad (9)$$

where $q_1 = \mu + \sigma Z_{(Q+1)}$, $q_3 = \mu + \sigma Z_{(3Q+1)}$. The detailed derivation of (9) is provided in the proof of Theorem 2 in Appendix 2. The numerical values of $w_{\text{opt}}(n)$ for $n$ from 5 to 101 are displayed in the middle panel of Figure 1. It is evident that $w_{\text{opt}}(n)$ is a decreasing function of $n$ with a lower bound around 0.7. We have further demonstrated in Theorem 2 in Appendix 2 that, from a theoretical point of view, the limit of $w_{\text{opt}}(n)$ is about 0.699 as $n$ tends to infinity. In contrast to the extreme values $a$ and $b$ in Scenario $\mathcal{S}_1$, the first and third quartiles are robust statistics and are equally important as the median in the estimation of the sample mean.

Noting that the optimal weight in (9) is rather complicated for practitioners, we therefore propose an approximation formula for $w_{\text{opt}}(n)$. In view of the middle panel of Figure 1 and also the theoretical limit in Theorem 2, we take the baseline to be 0.7 and approximate the remaining part to be a power function. That is, we consider the approximation form as $0.7 + c_1 n^{c_2}$. Finally, using the observed true weights, we figure out that the best coefficients are about $c_1 = 0.39$ and $c_2 = -1$. This leads to the approximated optimal weight as

$$\tilde{w}_{\text{opt}}(n) \approx 0.7 + \frac{0.39}{n} \qquad (10)$$

For researchers who prefer to use 0.699 as the baseline, the approximation formula is given as $\tilde{w}_{\text{opt}}(n) \approx 0.699 + 0.4/n$. Its performance, however, is very similar to the approximated formula in (10). To assess the accuracy of the approximation, we also display the values of $\tilde{w}_{\text{opt}}(n)$ in the second graph of Figure 1. It is evident that the approximated optimal weights fit well the true optimal weights, in particular for the sample size ranging from 5 to 101.

Finally, by plugging the approximation formula (10) into the estimator (8), we propose the estimator for Scenario $\mathcal{S}_2$ as

$$\bar{X}(w) \approx \left(0.7 + \frac{0.39}{n}\right) \frac{q_1 + q_3}{2} + \left(0.3 - \frac{0.39}{n}\right) m \qquad (11)$$

The performance of (11) is evaluated in Appendix 6, together with its numerical comparison with the estimation method in Wan et al.[4]

### 2.2.3　Improved estimation of the sample mean in $\mathcal{S}_3 = \{a, q_1, m, q_3, b; n\}$

For scenario $\mathcal{S}_3$, following the same spirit, we propose to estimate the sample mean by

$$\bar{X}(w_1, w_2) = w_1\left(\frac{a+b}{2}\right) + w_2\left(\frac{q_1 + q_3}{2}\right) + (1 - w_1 - w_2)m \qquad (12)$$

where $w_1$, $w_2$, and $(1 - w_1 - w_2)$ are the weights assigned to the mid-range $(a+b)/2$, the mid-quartile range $(q_1 + q_3)/2$, and the median $m$, respectively. Taking $w_1 = 0.25$ and $w_2 = 0.5$, the proposed estimator reduces to the estimator (3) proposed by Bland.[5] Theorem 3 in Appendix 2 shows that (12) is an unbiased estimator of $\mu$. Further, by minimizing the first-order derivatives of $\text{MSE}(\bar{X}(w_1, w_2))$, the optimal weights of $w_1$ and $w_2$ are given as,

$$\begin{pmatrix} w_{1,\text{opt}} \\ w_{2,\text{opt}} \end{pmatrix} = \begin{pmatrix} A + 4C - 4E & 4C + D - 2E - 2F \\ 4C + D - 2E - 2F & B + 4C - 4F \end{pmatrix}^{-1} \begin{pmatrix} 4C - 2E \\ 4C - 2F \end{pmatrix} \qquad (13)$$

where $A = \text{Var}(a+b)$, $B = \text{Var}(q_1 + q_3)$, $C = \text{Var}(m)$, $D = \text{Cov}(a+b, q_1+q_3)$, $E = \text{Cov}(a+b, m)$, and $F = \text{Cov}(q_1 + q_3, m)$.

To explore the behavior of the optimal weights, we plot the true values of $w_{1,\text{opt}}(n)$ and $w_{2,\text{opt}}(n)$ in the bottom panel of Figure 1. From the figure, we note that $w_{1,\text{opt}}(n)$ (the green solid points) is a decreasing function of $n$ with lower bound 0, and $w_{2,\text{opt}}(n)$ (the purple solid triangles) is an increasing function of $n$ with upper bound about 0.7. It is also noteworthy that $w_{1,\text{opt}}(n)$ and $w_{2,\text{opt}}(n)$ are both 0.4 when $n=5$. From the statistical point of view, when $n = 5$, the five-number summary is provided as the whole sample, and consequently, the sample mean which assigns a weight of 0.2 to each sample is the best unbiased estimator of $\mu$. This leads to $w_{1,opt}(n) = w_{2,opt}(n) = 0.4$. Note also that (13) is rather complicated for practical use. Following the similar structures as in (6) and (10), we propose to estimate the two optimal weights by $c_1/(c_1 + n^{c_2})$ and $0.7 - c_3 n^{c_4}$, respectively. Then by the observed true weights, the best values of the coefficients are $c_1 = 2.2$, $c_2 = 0.75$, $c_3 = 0.72$, and $c_4 = 0.55$. This leads to the approximated optimal weights as

$$\tilde{w}_{1,\text{opt}}(n) \approx \frac{2.2}{2.2 + n^{0.75}} \quad \text{and} \quad \tilde{w}_{2,\text{opt}}(n) \approx 0.7 - \frac{0.72}{n^{0.55}} \tag{14}$$

To assess the accuracy of the approximation, we also display the values of $\tilde{w}_{1,\text{opt}}(n)$ (the red line) and $\tilde{w}_{2,\text{opt}}(n)$ (the blue line) in the bottom panel of Figure 1. It is evident that the approximated optimal weights match precisely their respective values of the true optimal weights.

Finally, by plugging the approximation formula (14) into the estimator (12), we propose the estimator for scenario $S_3$ as

$$\bar{X}(w_1, w_2) \approx \left(\frac{2.2}{2.2 + n^{0.75}}\right)\frac{a+b}{2} + \left(0.7 - \frac{0.72}{n^{0.55}}\right)\frac{q_1 + q_3}{2} + \left(0.3 + \frac{0.72}{n^{0.55}} - \frac{2.2}{2.2 + n^{0.75}}\right)m \tag{15}$$

The performance of (15) is evaluated in Appendix 7, together with its numerical comparison with the estimation method in Bland.[5]

## 3 Simulation studies

To compare the performance between existing methods and our newly proposed methods, we conduct some simulation studies. Using the same settings as in Hozo et al.,[3] five different distributions are taken into consideration: the normal distribution with mean $\mu = 50$ and standard deviation $\sigma = 17$, the log-normal distribution with location parameter $\mu = 4$ and scale parameter $\sigma = 0.3$, the beta distribution with shape parameters $\alpha = 9$ and $\beta = 4$, the exponential distribution with rate parameter $\lambda = 10$, and the Weibull distribution with shape parameter $k = 2$ and scale parameter $\lambda = 35$.

For the $i$th simulation, we generate a random sample of $n$ observations from a specified distribution and compute the sample mean $\bar{X}_i$. We also compute the sample mean from the median, minimum, and maximum values and/or the first and third quartiles by using the existing and new methods, denoted by $\bar{X}_i^{\text{Ex}}$ and $\bar{X}_i^{\text{New}}$, respectively. To evaluate the performance of the proposed "rules of thumb," we then compute the relative mean squared error (RMSE) of the estimators as

$$\text{RMSE}(\bar{X}^{\text{Ex}}) = \frac{\sum_{i=1}^{T}(\bar{X}_i^{\text{Ex}} - \mu)^2}{\sum_{i=1}^{T}(\bar{X}_i - \mu)^2} \quad \text{and} \quad \text{RMSE}(\bar{X}^{\text{New}}) = \frac{\sum_{i=1}^{T}(\bar{X}_i^{\text{New}} - \mu)^2}{\sum_{i=1}^{T}(\bar{X}_i - \mu)^2} \quad (16)$$

where $\mu$ is the true mean value and $T$ is the total number of repetitions. The smaller the RMSE is, the better accuracy is achieved. It is also noteworthy that the lower bound of RMSE is 1, in which the approximated mean estimation performs equally well as the true sample mean. Moreover, to save space, we only provide the results of the simulation study for scenario $\mathcal{S}_1$. The other two simulation studies will be provided as Appendices 6 and 7.

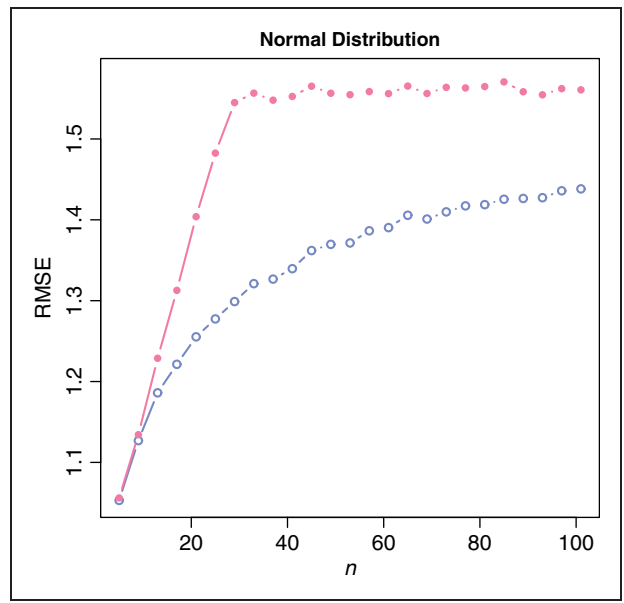

**Figure 2.** RMSE of the sample mean estimation for data from normal distribution for scenario $S_1$. The pink line with solid circles represents Hozo et al.'s method, and the light blue line with empty circles represents the new estimator.

In this simulation study, we compare Hozo et al.'s estimator (1) and our proposed estimator (7). Figure 2 reports the RMSE of 100,000 simulations for normal distribution with the sample size ranging from 5 to 101. It is obvious that our new method has a much smaller RMSE than Hozo et al.'s method. When $n$ increases, the two methods tend to have a similar performance as the optimal weight of $(a + b)/2$ reduces to about zero for large sample sizes. Figure 3 provides the simulation results of the other four distributions as we mentioned before. We observe that for each of skewed distributions, our new estimator provides a more accurate estimate of the true mean than Hozo et al.'s method, especially when $n$ is large. Note also that Hozo et al.'s method has a change point near $n=25$ (as suggested in (1)), especially for log-normal distribution and exponential distribution. In conclusion, we have provided an optimal and smoothly changed weight for $(a+b)/2$, or equivalently for $m$, which makes the new estimator more adaptive and more stable than Hozo et al.'s method for scenario $\mathcal{S}_1$, no matter whether the data are normal or skewed.

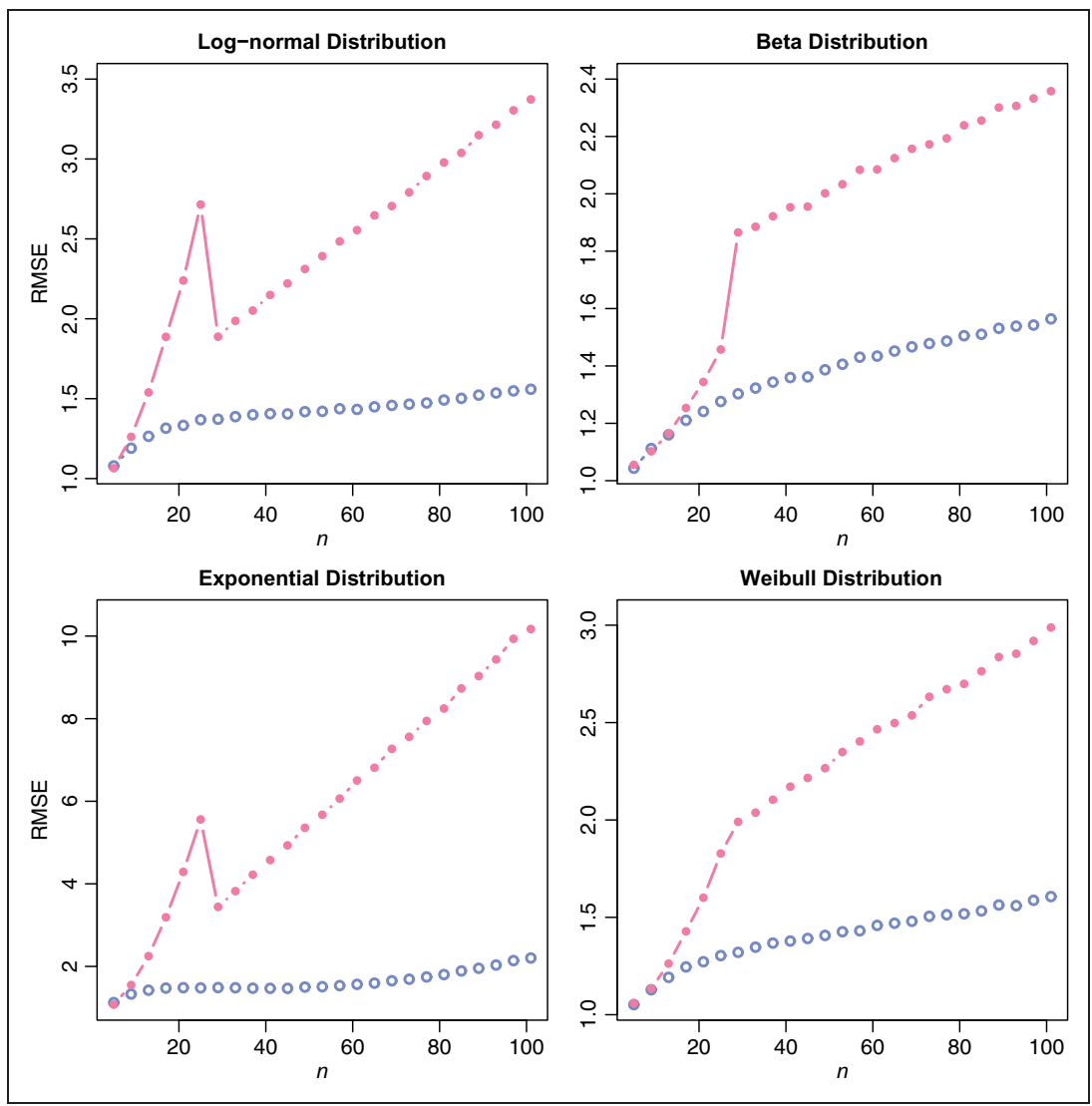

**Figure 3.** RMSE of the sample mean estimation for data from non-normal distributions for scenario $\mathcal{S}_1$. The pink line with solid circles represents Hozo et al.'s method, and the light blue line with empty circles represents the new estimator.

## 4 Real data analysis

To illustrate the potential value of our method in real data analysis, we collect some real data and compare the estimations using our methods with the ones using the existing method. The collected data are from a systematic review and meta-analysis of the association between low serum vitamin D and risk of active tuberculosis in humans.[8]

### 4.1 Data description

In Nnoaham and Clarke,[8] the summary statistics reported from seven studies were used to conduct the meta-analysis. Among those seven studies, three of them only reported the sample median and range, which is exactly the case of Scenario $\mathcal{S}_1$. For these three studies, the sample mean and standard deviation need to be estimated from the sample median and range in order to calculate the pooled effect size. The summary statistics are presented in Table 1, in which Studies 1 to 3 reported the median, minimum, and maximum values, Studies 4 and 5 reported the mean values and standard deviations, Study 6 reported the odds ratio for vitamin D deficiency in tuberculosis cases compared with controls, and Study 7 reported the mean value and range (i.e. difference between the minimum and maximum).

### 4.2 Results and comparison

To conduct a random effect meta-analysis, Nnoaham and Clarke first used Hozo et al.'s method to estimate the sample mean and standard deviation for the first three studies and the sample standard deviation for Study 7 in Table 1. Next, the mean difference (i.e. the Cohen's $d$ value[9]) is computed as the effect size. The odds ratio in Study 6 is directly converted to the effect size by Chinn.[10] Finally, the pooled effect size is computed and the heterogeneity between studies will be assessed using the $\chi^2$ statistic and the $I^2$ index (the amount of variation due to heterogeneity). Their results are presented in Table 2. It is worth to mention that in Nnoaham and Clarke,[8] they mistakenly reported the estimated effect size of Study 2.

Table 1. Summary of included studies.

| Index | Study | Size (cases) | Size (controls) | Results (serum Vitamin D levels) |
|---|---|---|---|---|
| 1 | Davies et al.[11] | 40 | 40 | Median (range) in: Cases 16.0 nmol/L (2.25–74.25 nmol/L), Controls 27.25 nmol/L (9.0–132.5 nmol/L) |
| 2 | Grange et al.[12] | 40 | 38 | Median (range) in: Cases 65.75 nmol/L (43.75–130.5 nmol/L), Controls 69.5 nmol/L (48.5–125 nmol/L) |
| 3 | Davies et al.[13] | 15 | 15 | Median (range) in: Cases 39.75 nmol/L (16.75–89.25 nmol/L), Controls 65.5 nmol/L (26.25–114.75 nmol/L) |
| 4 | Davies et al.[14] | 51 | 51 | Mean (SD) in: Cases 69.5 nmol/L (24.5 nmol/L), Controls 95.5 nmol/L (29.25 nmol/L) |
| 5 | Chan et al.[15] | 24 | 24 | Mean (SD) in: Cases 46.5 nmol/L (18.5 nmol/L), Controls 52.25 nmol/L (15.75 nmol/L) |
| 6 | Wilkinson et al.[16] | 103 | 42 | Odds ratio (CI) of cases compared with controls 2.9 (1.3–6.5) |
| 7 | Sasidharan et al.[17] | 35 | 16 | Mean (range) in: Cases 26.75 nmol/L (2.5–75 nmol/L), Controls 48.5 nmol/L (22.5–145 nmol/L) |

Following the same aforementioned procedure, we use equation (7) to estimate the sample mean for the first studies and the method of Wan et al.[4] to estimate the sample standard deviation for the first three studies and Study 7. Then, we further use the same method as Nnoaham and Clarke did to compute the pooled effect size, the $\chi^2$ statistic, and the $I^2$ index with our estimations. The new results are reported in Table 3.

Comparing Tables 2 and 3, we can observe some significant differences between the old results and the new results. The most noticeable one is the effect size difference for Study 1. The effect size using Hozo et al.'s method is within the range of large effect level while it only reaches the median effect level using our method (i.e. 0.8656 vs. 0.6622). The effect size different for Studies 2, 3, and 7 is also non-trivial. Although the pooled effect size from both methods seems to be close to each other, we found that the $I^2$ indices for heterogeneity between studies are quite different (i.e. 48.54% from Hozo et al.'s method and 34.85% from our method). According to Higgins et al.,[18] the value 48.54% of $I^2$ is very close to moderate heterogeneity level while the value 34.85% is close to little heterogeneity level. It is obvious that using our method in this study may eventually lead to a different conclusion.

**Table 2.** Effect sizes of low serum vitamin D in tuberculosis (using Hozo et al.'s method).

| Index | Study | Size (cases) | Size (control) | Effect size (SE) | Weight | 95% CI of effect size |
|---|---|---|---|---|---|---|
| 1 | Davies et al.[11] | 40 | 40 | 0.8656 (0.0562) | 17.79 | [0.4043, 1.3218] |
| 2 | Grange et al.[12] | 40 | 38 | 0.0824 (0.0527) | 18.97 | [−0.3621, 0.5263] |
| 3 | Davies et al.[13] | 15 | 15 | 0.9190 (0.1590) | 6.29 | [0.1569, 1.6664] |
| 4 | Davies et al.[14] | 51 | 51 | 0.9637 (0.0447) | 22.35 | [0.5511, 1.3719] |
| 5 | Chan et al.[15] | 22 | 23 | 0.3353 (0.0944) | 10.59 | [−0.2554, 0.9221] |
| 6 | Wilkinson et al.[16] | 103 | 42 | 0.5882 (0.0352) | 28.40 | [0.2220, 0.9525] |
| 7 | Sasidharan et al.[17] | 35 | 16 | 0.9584 (0.1045) | 9.57 | [0.3329, 1.5749] |
| | Total (95% CI) | 306 | 225 | 0.6732 | 100.00 | [0.4961, 0.8498] |
| | Test for heterogeneity: $\chi^2 = 11.6594, df = 6 (P = 0.07), I^2 = 48.539\%$ | | | | | |

**Table 3.** Effect sizes of low serum vitamin D in tuberculosis (using the new method).

| Index | Study | Size (cases) | Size (control) | Effect size (SE) | Weight | 95% CI of effect Size |
|---|---|---|---|---|---|---|
| 1 | Davies et al.[11] | 40 | 40 | 0.6622 (0.0542) | 18.46 | [0.2098, 1.1105] |
| 2 | Grange et al.[12] | 40 | 38 | 0.1588 (0.0528) | 18.93 | [−0.2864, 0.6030] |
| 3 | Davies et al.[13] | 15 | 15 | 0.9852 (0.1614) | 6.19 | [0.2171, 1.7378] |
| 4 | Davies et al.[14] | 51 | 51 | 0.9637 (0.0447) | 22.35 | [0.5511, 1.3719] |
| 5 | Chan et al.[15] | 22 | 23 | 0.3353 (0.0944) | 10.59 | [−0.2554, 0.9221] |
| 6 | Wilkinson et al.[16] | 103 | 42 | 0.5882 (0.0352) | 28.40 | [0.2220, 0.9525] |
| 7 | Sasidharan et al.[17] | 35 | 16 | 0.9084 (0.1036) | 9.66 | [0.2861, 1.5223] |
| | Total (95% CI) | 306 | 225 | 0.6257 | 100.00 | [0.4510, 0.8035] |
| | Test for heterogeneity: $\chi^2 = 9.2091, df = 6 (P = 0.162), I^2 = 34.847\%$ | | | | | |

**Table 4.** Summary table for estimating $\bar{X}$ under different scenarios.

| | Scenario $\mathcal{S}_1$ | Scenario $\mathcal{S}_2$ | Scenario $\mathcal{S}_3$ |
|---|---|---|---|
| Hozo et al.[3] | Equation (1) | – | – |
| Wan et al.[4] | – | Equation (2) | – |
| Bland[5] | – | – | Equation (3) |
| New methods | Equation (7) | Equation (11) | Equation (15) |

## 5 Conclusion

Meta-analysis is a popular method in evidence-based medicine to provide an overall estimation of a treatment effectiveness from a set of similar clinical trials. The sample mean and standard deviation are often used in meta-analysis but sometimes the results are recorded using the median, the minimum and maximum values, and/or the first and third quartiles. Searching for a reliable approximation method to obtain the sample mean and standard deviation and for conducting further research has emerged as a popular topic. The estimation of the sample standard deviation has been thoroughly discussed and significantly improved in Wan et al.[4] But the current estimation of the sample mean adopts either the famous method proposed by Hozo et al.[3] or its extension by Bland.[5] One major limitation of such methods, however, is that the information of the sample size is not fully used or even ignored in the sample mean estimation.

For the three frequently encountered scenarios, the simulation studies show that our newly proposed methods, which incorporate the sample size via a smoothly changing weight in the estimation, greatly improve the existing methods. For all scenarios, we provide both theoretical and empirical computations for the optimal weights. The simulation results show that the empirical computation of optimal weight not only matches the theoretical computation with high accuracy but has almost the same simplicity as the existing methods. Here, we provide a summary table of the new estimators of the sample mean in different scenarios, which may serve as a comprehensive guidance for researchers when performing meta-analysis. To help the researchers to utilize the proposed mean estimators, an Excel spread sheet containing all estimators in Table 4 is provided as the additional file and is freely available at www.math.hkbu.edu.hk/~tongt/papers/optimalmean.xlsx. Using the spread sheet, users can easily obtain the sample mean values by providing the corresponding information for appropriate scenario such as the sample size, median and extremum values. We also provide the formulas for Hozo et al.'s, Bland's and Wan et al.'s methods in the Excel spread sheet for the comparison. To further illustrate the performance of the newly proposed methods, a real meta-analysis was conducted using seven studies from a systematic review and meta-analysis of the association between low serum vitamin D and risk of active tuberculosis in humans.[8] We compared the effect sizes obtained from our proposed methods with those from the existing methods. It is evident that there are some significant differences between the new results and the old ones. Since the simulation studies indicate that the new methods could improve the estimation performance, we expect the proposed estimators may help researchers to make more convincing conclusions when conducting meta-analysis in real-world settings.

### Acknowledgements

The authors thank the editor, the associate editor, and two reviewers for their constructive comments that led to a substantial improvement of the article.

### Declaration of Conflicting Interests

The author(s) declared no potential conflicts of interest with respect to the research, authorship, and/or publication of this article.

### Funding

The author(s) disclosed receipt of the following financial support for the research, authorship, and/or publication of this article: Xiang Wan's research was supported by the HKBU grant FRG2/14-15/077 and the Hong Kong RGC grant HKBU12202114. Tiejun Tong's research was supported by the HKBU grants FRG1/14-15/084 and FRG2/15-16/019, and the HKBU Century Club Sponsorship Scheme in 2016.

## Appendix 1

### Some preliminary results

To derive the optimal weights for the three scenarios, we first present some preliminary results for the normal distribution and for the associated order statistics. The normal distribution $N(\mu, \sigma^2)$ is commonly used in statistics for data analysis. Its probability density function (PDF) is given as

$$\phi(x|\mu, \sigma^2) = \frac{1}{\sqrt{2\pi\sigma^2}} \exp\left\{-\frac{(x-\mu)^2}{2\sigma^2}\right\}$$

where $\mu$ is the mean value and $\sigma^2$ is the variance, or equivalently, $\sigma$ is the standard deviation. For the normal distribution, $\mu$ is also known as the median and the mode. When $\mu = 0$ and $\sigma^2 = 1$, the distribution reduces to the standard normal distribution $N(0, 1)$. Let also $\Phi(\cdot)$ be the cumulative density function (CDF) of the standard normal distribution. By symmetry, we have $\phi(z) = \phi(-z)$ and $\Phi(z) = 1 - \Phi(-z)$.

To investigate the properties of the five-number summary for the data, we introduce some theoretical results for the order statistics $Z_{(1)} \leq \cdots \leq Z_{(n)}$ of the random sample $\{Z_1, \ldots, Z_n\}$ from the standard normal distribution. By symmetry, $Z_{(i)}$ and $-Z_{(n-i+1)}$ follow the same distribution, and $(Z_{(i)}, Z_{(j)})$ and $(Z_{(n-i+1)}, Z_{(n-j+1)})$ follow the same joint distribution. According to Arnold and Balakrishnan,[19] Chen,[20] and Ahsanullah et al.,[21] we have the following two lemmas.

### Lemma 1

Let $Z_1, \ldots, Z_n$ be a random sample of $N(0, 1)$, and $Z_{(1)} \leq \cdots \leq Z_{(n)}$ be the ordered statistics $Z_1, \ldots, Z_n$. Then

$$E(Z_{(i)}) = -E(Z_{(n-i+1)}), \quad 1 \leq i \leq n,$$
$$E(Z_{(i)}Z_{(j)}) = E(Z_{(n-i+1)}Z_{(n-j+1)}), \quad 1 \leq i \leq j \leq n$$

### Lemma 2

Let $Z_1, \ldots, Z_n$ be a random sample of $N(0, 1)$, and $Z_{[np]}$ be the $p$th quantile of the sample, where $[np]$ denotes the integer part of $np$. Let also $\Phi^{-1}(\cdot)$ be the inverse function of $\Phi(\cdot)$.

(i) For any $0 < p < 1$, we have

$$\sqrt{n}(Z_{[np]} - \Phi^{-1}(p)) \xrightarrow{d} N\left(0, \frac{p(1-p)}{[\phi(\Phi^{-1}(p))]^2}\right), \quad \text{as } n \to \infty$$

where $\xrightarrow{d}$ denotes convergence in distribution.

(ii) For any $0 < p_1 < p_2 < 1$, as $n \to \infty$, $(Z_{[np_1]}, Z_{[np_2]})$ follows asymptotically a bivariate normal distribution with mean vector $(\Phi^{-1}(p_1), \Phi^{-1}(p_2))$ and covariance matrix $\Sigma = (\sigma_{ij})_{2 \times 2}$, where $\sigma_{12} = \sigma_{21}$ and

$$\sigma_{ij} = \frac{p_i(1-p_j)}{n\phi(\Phi^{-1}(p_i))\phi(\Phi^{-1}(p_j))}, \quad 1 \leq i \leq j \leq 2$$

## Appendix 2
## Theoretical results of the proposed estimators

Recall that in "Improved methods" section, $X_1, \ldots, X_n$ are defined as a random sample of size $n$ from the normal distribution $N(\mu, \sigma^2)$, and $X_{(1)} \leq \cdots \leq X_{(n)}$ are the ordered statistics of the sample. Meanwhile, they can represented as $X_i = \mu + \sigma Z_i$ and $X_{(i)} = \mu + \sigma Z_{(i)}$ for $i = 1, \ldots, n$. By the two lemmas in Appendix 1, we have the following theoretical results for the proposed estimators under three scenarios, respectively.

**Theorem 1**

For the estimator $\bar{X}(w)$ in (4) for scenario $\mathcal{S}_1 = \{a, m, b; n\}$, i.e.

$$\bar{X}(w) = w\left(\frac{a+b}{2}\right) + (1-w)m$$

we have the following conclusions:

(i) $\bar{X}(w)$ in (4) is an unbiased estimator of $\mu$, i.e. $E(\bar{X}(w)) = \mu$.
(ii) $\text{MSE}(\bar{X}(w)) = (w^2/4)\text{Var}(a+b) + (1-w)^2\text{Var}(m) + w(1-w)\text{Cov}(a+b, m)$.
(iii) $w_{\text{opt}}(n) = [4\text{Var}(m) - 2\text{Cov}(a+b, m)]/[\text{Var}(a+b) + 4\text{Var}(m) - 4\text{Cov}(a+b, m)]$.

**Proof**

(i) The expected value of the proposed estimator is

$$\begin{aligned}
E(\bar{X}(w)) &= \frac{w}{2}E(a) + \frac{w}{2}E(b) + (1-w)E(m) \\
&= \frac{w}{2}E(\mu + \sigma Z_{(1)}) + \frac{w}{2}E(\mu + \sigma Z_{(n)}) + (1-w)E(\mu + \sigma Z_{(2Q+1)}) \\
&= \frac{w}{2}(\mu + \sigma E(Z_{(1)})) + \frac{w}{2}(\mu + \sigma E(Z_{(n)})) + (1-w)(\mu + \sigma E(Z_{(2Q+1)})) \\
&= \mu + \sigma[E(Z_{(1)}) + E(Z_n)] + \sigma E(Z_{(2Q+1)})
\end{aligned}$$

By Lemma 1 with $i = 1$, we have $E(Z_{(1)}) = -E(Z_{(n)})$; similarly with $i = 2Q + 1$, we have $E(Z_{(2Q+1)}) = 0$. This shows that $\bar{X}(w)$ is an unbiased estimator of $\mu$.

(ii) By the result in (i), we have $\text{Bias}(\bar{X}(w)) = 0$. Then

$$\begin{aligned}
\text{MSE}(\bar{X}(w)) &= \text{Var}(\bar{X}(w)) \\
&= \text{Var}\left[\frac{w}{2}(a+b) + (1-w)m\right] \\
&= \frac{w^2}{4}\text{Var}(a+b) + (1-w)^2\text{Var}(m) + w(1-w)\text{Cov}(a+b, m)
\end{aligned}$$

(iii) The first derivative of MSE with respect to $w$ is

$$\frac{d}{dw}\text{MSE}(\bar{X}(w)) = \frac{w}{2}\text{Var}(a+b) + 2(w-1)\text{Var}(m) + (1-2w)\text{Cov}(a+b,m)$$

Letting the first derivative equal to zero, we have the solution of $w$ as

$$w = \frac{4\text{Var}(m) - 2\text{Cov}(a+b,m)}{\text{Var}(a+b) + 4\text{Var}(m) - 4\text{Cov}(a+b,m)} \qquad (17)$$

Further by Cauchy–Schwarz inequality,

$$\frac{d^2}{dw^2}\text{MSE}(\bar{X}(w)) = \frac{1}{2}\text{Var}(a+b) + 2\text{Var}(m) - 2\text{Cov}(a+b,m)$$
$$= \frac{1}{2}[\text{Var}(a+b) + \text{Var}(2m) - 2\text{Cov}(a+b,2m)] \geq 0,$$

we conclude that the derived $w$ in (17) is the optimal weight for the proposed estimator.

**Theorem 2**

For Scenario $\mathcal{S}_2 = \{q_1, m, q_3; n\}$, recall the estimator $\bar{X}(w)$ in (8), i.e.

$$\bar{X}(w) = w\left(\frac{q_1+q_3}{2}\right) + (1-w)m$$

we have the following conclusions:

(i) $\bar{X}(w)$ in (8) is an unbiased estimator of $\mu$, i.e. $E(\bar{X}(w)) = \mu$.
(ii) $\text{MSE}(\bar{X}(w)) = (w^2/4)\text{Var}(q_1+q_3) + (1-w)^2\text{Var}(m) + w(1-w)\text{Cov}(q_1+q_3,m)$.
(iii) $w_{\text{opt}}(n) = [4\text{Var}(m) - 2\text{Cov}(q_1+q_3,m)]/[\text{Var}(q_1+q_3) + 4\text{Var}(m) - 4\text{Cov}(q_1+q_3,m)]$.
(iv) When $n$ is large, $w_{\text{opt}}(n) \approx 0.699$.

**Proof**

(i) Following the same procedure as proving (i) of Theorem 1, we need to compute $E(\bar{X}(w))$,

$$E(\bar{X}(w)) = \frac{w}{2}E(q_1) + \frac{w}{2}E(q_3) + (1-w)E(m)$$
$$= \frac{w}{2}E(X_{(Q+1)}) + \frac{w}{2}E(X_{(3Q+1)}) + (1-w)E(X_{(2Q+1)})$$
$$= \frac{w}{2}E(\mu + \sigma Z_{(Q+1)}) + \frac{w}{2}E(\mu + \sigma Z_{(3Q+1)}) + (1-w)E(\mu + \sigma Z_{(2Q+1)})$$
$$= \frac{w}{2}(\mu + \sigma E(Z_{(Q+1)})) + \frac{w}{2}(\mu + \sigma E(Z_{(3Q+1)})) + (1-w)(\mu + \sigma E(Z_{(2Q+1)}))$$
$$= w\mu + (1-w)\mu + \sigma[E(Z_{(Q+1)}) + E(Z_{(3Q+1)})] + \sigma E(Z_{(2Q+1)}).$$

By using Lemma 1, when $i = Q+1$ or $3Q+1$, we have $E(Z_{(Q+1)}) = -E(Z_{(3Q+1)})$. By proving (i) of Theorem 1, we know that $E(Z_{(2Q+1)}) = 0$. This indicates that $\bar{X}(w)$ is an unbiased estimator of $\mu$.

(ii) By (i), we have $\text{Bias}(\bar{X}(w)) = 0$, then the MSE of the estimator $\bar{X}(w)$ is,

$$\text{MSE}(\bar{X}(w)) = \text{Var}(\bar{X}(w)) = \text{Var}\left[\frac{w}{2}(q_1+q_3) + (1-w)m\right]$$
$$= \frac{w^2}{4}\text{Var}(q_1+q_3) + (1-w)^2\text{Var}(m) + w(1-w)\text{Cov}(q_1+q_3,m)$$

(iii) The first derivative of MSE with respect to $w$ is

$$\frac{d}{dw}\text{MSE}(\bar{X}(w)) = \frac{w}{2}\text{Var}(q_1+q_3) + 2(w-1)\text{Var}(m) + (1-2w)\text{Cov}(q_1+q_3,m)$$

Letting the first derivative equal to zero, we have the solution of $w$ as

$$w = \frac{4\text{Var}(m) - 2\text{Cov}(q_1 + q_3, m)}{\text{Var}(q_1 + q_3) + 4\text{Var}(m) - 4\text{Cov}(q_1 + q_3, m)} \tag{18}$$

By Cauchy–Schwarz inequality,

$$\frac{d^2}{dw^2}\text{MSE}(\bar{X}(w)) = \frac{1}{2}\text{Var}(q_1 + q_3) + 2\text{Var}(m) - 2\text{Cov}(q_1 + q_3, m)$$
$$= \frac{1}{2}[\text{Var}(q_1 + q_3) + \text{Var}(2m) - 2\text{Cov}(q_1 + q_3, 2m)] \geq 0,$$

we conclude that the derived $w$ in (18) is the optimal weight for the proposed estimator.

(iv) Let $p_1 = 0.25$, $p_2 = 0.5$ and $p_3 = 0.75$, the sample first and third quartiles and the median are then be represented by $q_1 = \mu + \sigma Z_{[0.25n]}$, $m = \mu + \sigma Z_{[0.5n]}$ and $q_3 = \mu + \sigma Z_{[0.75n]}$. By Lemma 2, when $n$ is large, we have

$$\begin{aligned}
\text{Var}(Z_{[0.25n]}) &= \frac{p_1(1 - p_1)}{n[\phi(\Phi^{-1}(p_1))]^2} = \frac{0.25(0.75)}{n[\phi(\Phi^{-1}(0.25))]^2} \approx \frac{1.8568}{n}, \\
\text{Var}(Z_{[0.5n]}) &= \frac{0.5^2}{n[\phi(\Phi^{-1}(0.5))]^2} = \frac{\pi}{2n}, \\
\text{Var}(Z_{[0.75n]}) &= \frac{0.25(0.75)}{n[\phi(\Phi^{-1}(0.75))]^2} \approx \frac{1.8568}{n}, \\
\text{Cov}(Z_{[0.25n]}, Z_{[0.5n]}) &= \frac{0.25(0.5)}{n\phi(\Phi^{-1}(0.25))\phi(\Phi^{-1}(0.5))} \approx \frac{0.9860}{n}, \\
\text{Cov}(Z_{[0.25n]}, Z_{[0.75n]}) &= \frac{0.25^2}{n\phi(\Phi^{-1}(0.25))\phi(\Phi^{-1}(0.75))} \approx \frac{0.6189}{n}
\end{aligned} \tag{19}$$

Hence, we could easily obtain that $\text{Var}(q_1) = \text{Var}(q_3) \approx 1.8568\sigma^2/n$, $\text{Var}(m) \approx \pi\sigma^2/2n$, $\text{Cov}(q_1, m) \approx 0.9860\sigma^2/n$ and $\text{Cov}(q_1, q_3) \approx 0.6189\sigma^2/n$, as $n \to \infty$.

By (iii), plugging in the above information into the optimal weight formula (18),

$$\begin{aligned}
w_{\text{opt}}(n) &= \frac{4\text{Var}(m) - 4\text{Cov}(q_1, m)}{\text{Var}(q_1) + \text{Var}(q_3) + \text{Cov}(q_1, q_3) + 4\text{Var}(m) - 8\text{Cov}(q_1, m)} \\
&\approx \frac{2\pi - 4(0.9860)}{2(1.8568) + 2(0.6189) + 2\pi - 8(0.9860)} \approx 0.699
\end{aligned}$$

As a result, for the approximation model of $w_{\text{opt}}(n)$, it is reasonable to choose 0.7 as its baseline.

**Theorem 3**
For Scenario $S_3 = \{a, q_1, m, q_3, b; n\}$, recall the estimator $\bar{X}(w)$ in (12), i.e.

$$\bar{X}(w) = w_1\left(\frac{a+b}{2}\right) + w_2\left(\frac{q_1+q_3}{2}\right) + (1 - w_1 - w_2)m$$

we have the following conclusions:

(i) $\bar{X}(w)$ in (12) is an unbiased estimator of $\mu$, i.e. $E(\bar{X}(w)) = \mu$.
(ii) The MSE of the estimator is given as

$$\begin{aligned}
\text{MSE}(\bar{X}(w_1, w_2)) = &(w_1^2/4)\text{Var}(a + b) + (w_2^2/4)\text{Var}(q_1 + q_3) + (1 - w_1 - w_2)^2\text{Var}(m) \\
&+ (w_1 w_2/2)\text{Cov}(a + b, q_1 + q_3) + w_1(1 - w_1 - w_2)\text{Cov}(a + b, m) \\
&+ w_2(1 - w_1 - w_2)\text{Cov}(q_1 + q_3, m)
\end{aligned}$$

(iii) The optimal weights of $(a+b)/2$ and $(q_1+q_3)/2$ are

$$\begin{pmatrix} w_{1,\text{opt}} \\ w_{2,\text{opt}} \end{pmatrix} = \begin{pmatrix} A+4C-4E & 4C+D-2E-2F \\ 4C+D-2E-2F & B+4C-4F \end{pmatrix}^{-1} \begin{pmatrix} 4C-2E \\ 4C-2F \end{pmatrix}$$

**Proof**
(i) The expected value of the estimator can be obtained as

$$\begin{aligned} E(\bar{X}(w_1,w_2)) &= \frac{w_1}{2}E(a) + \frac{w_1}{2}E(b) + \frac{w_2}{2}E(q_1) + \frac{w_2}{2}E(q_3) + (1-w_1-w_2)E(m) \\ &= \frac{w_1}{2}E(\mu+\sigma Z_{(1)}) + \frac{w_1}{2}E(\mu+\sigma Z_{(n)}) + \frac{w_2}{2}E(\mu+\sigma Z_{(Q+1)}) \\ &\quad + \frac{w_2}{2}E(\mu+\sigma Z_{(3Q+1)}) + (1-w_1-w_2)E(\mu+\sigma Z_{(2Q+1)}) \\ &= \mu + \frac{w_1}{2}\sigma[E(Z_{(1)}) + E(Z_{(n)})] + \frac{w_2}{2}\sigma[E(Z_{(Q+1)}) + E(Z_{3Q+1})] + \sigma E(Z_{(2Q+1)}) \end{aligned}$$

By Lemma 1, with $i=1, Q+1, 2Q+1, 3Q+1$ or $n$, we have $E(Z_{(1)}) = -E(Z_{(n)})$, $E(Z_{(Q+1)}) = -E(Z_{(3Q+1)})$ and $E(Z_{(2Q+1)}) = 0$, which shows that $\bar{X}(w)$ is an unbiased estimator of $\mu$.

(ii) By (i), we have $\text{Bias}(\bar{X}(w_1,w_2)) = 0$. Then,

$$\begin{aligned} \text{MSE}(\bar{X}(w_1,w_2)) &= \text{Var}\left[\frac{w_1}{2}(a+b) + \frac{w_2}{2}(q_1+q_3) + (1-w_1-w_2)m\right] \\ &= \frac{w_1^2}{4}\text{Var}(a+b) + \frac{w_2^2}{4}\text{Var}(q_1+q_3) + (1-w_1-w_2)^2\text{Var}(m) \\ &\quad + \frac{w_1 w_2}{2}\text{Cov}(a+b,q_1+q_3) + w_1(1-w_1-w_2)\text{Cov}(a+b,m) \\ &\quad + w_2(1-w_1-w_2)\text{Cov}(q_1+q_3,m) \end{aligned}$$

(iii) With the definition of $A$ through $F$ in "Improved estimation of the sample mean in $\mathcal{S}_3 = \{a,q_1,m,q_3,b;n\}$" section, the two first partial derivatives of MSE with respect to $w_1$ and $w_2$ are

$$\begin{cases} \frac{\partial}{\partial w_1}\text{MSE}(\bar{X}(w_1,w_2)) = w_1(A+4C-4E) + w_2(4C+D-2E-2F) - (4C-2E), \\ \frac{\partial}{\partial w_2}\text{MSE}(\bar{X}(w_1,w_2)) = w_1(4C+D-2E-2F) + w_2(B+4C-4F) - (4C-2F) \end{cases}$$

Letting the first partial derivatives be zero, we have

$$\begin{pmatrix} A+4C-4E & 4C+D-2E-2F \\ 4C+D-2E-2F & B+4C-4F \end{pmatrix} \begin{pmatrix} w_1 \\ w_2 \end{pmatrix} = \begin{pmatrix} 4C-2E \\ 4C-2F \end{pmatrix} \quad (20)$$

This leads to

$$\begin{pmatrix} w_1 \\ w_2 \end{pmatrix} = \begin{pmatrix} A+4C-4E & 4C+D-2E-2F \\ 4C+D-2E-2F & B+4C-4F \end{pmatrix}^{-1} \begin{pmatrix} 4C-2E \\ 4C-2F \end{pmatrix} \quad (21)$$

Further, we can verify that the coefficient matrix in the left side of (20) is positive definite. Hence, $\text{MSE}(\bar{X}(w_1,w_2))$ is a convex function of $w_1$ and $w_2$. This shows that the derived $w_1$ and $w_2$ in (21) are the optimal weights for the proposed estimator.

# Appendix 3

## True optimal weights in scenario $\mathcal{S}_1$

**Table 5.** The true optimal weights for $(a+b)/2$ in scenario $\mathcal{S}_1$, with $n$ ranging from 5 to 501.

| n | $w_{\text{opt}}(n)$ | n | $w_{\text{opt}}(n)$ | n | $w_{\text{opt}}(n)$ | n | $w_{\text{opt}}(n)$ | n | $w_{\text{opt}}(n)$ |
|---|---|---|---|---|---|---|---|---|---|
| 5 | 0.5514 | 105 | 0.1084 | 205 | 0.0671 | 305 | 0.0497 | 405 | 0.0399 |
| 9 | 0.4346 | 109 | 0.1056 | 209 | 0.0661 | 309 | 0.0492 | 409 | 0.0396 |
| 13 | 0.3682 | 113 | 0.1030 | 213 | 0.0652 | 313 | 0.0487 | 413 | 0.0393 |
| 17 | 0.3232 | 117 | 0.1005 | 217 | 0.0643 | 317 | 0.0483 | 417 | 0.0391 |
| 21 | 0.2903 | 121 | 0.0982 | 221 | 0.0635 | 321 | 0.0477 | 421 | 0.0387 |
| 25 | 0.2642 | 125 | 0.0960 | 225 | 0.0626 | 325 | 0.0474 | 425 | 0.0384 |
| 29 | 0.2435 | 129 | 0.0939 | 229 | 0.0617 | 329 | 0.0468 | 429 | 0.0382 |
| 33 | 0.2263 | 133 | 0.0918 | 233 | 0.0609 | 333 | 0.0465 | 433 | 0.0379 |
| 37 | 0.2118 | 137 | 0.0900 | 237 | 0.0602 | 337 | 0.0460 | 437 | 0.0376 |
| 41 | 0.1992 | 141 | 0.0881 | 241 | 0.0595 | 341 | 0.0456 | 441 | 0.0373 |
| 45 | 0.1882 | 145 | 0.0863 | 245 | 0.0588 | 345 | 0.0452 | 445 | 0.0370 |
| 49 | 0.1786 | 149 | 0.0847 | 249 | 0.0580 | 349 | 0.0448 | 449 | 0.0368 |
| 53 | 0.1702 | 153 | 0.0830 | 253 | 0.0573 | 353 | 0.0445 | 453 | 0.0366 |
| 57 | 0.1626 | 157 | 0.0815 | 257 | 0.0566 | 357 | 0.0440 | 457 | 0.0362 |
| 61 | 0.1557 | 161 | 0.0800 | 261 | 0.0560 | 361 | 0.0436 | 461 | 0.0360 |
| 65 | 0.1495 | 165 | 0.0786 | 265 | 0.0554 | 365 | 0.0433 | 465 | 0.0358 |
| 69 | 0.1438 | 169 | 0.0773 | 269 | 0.0547 | 369 | 0.0429 | 469 | 0.0356 |
| 73 | 0.1385 | 173 | 0.0760 | 273 | 0.0541 | 373 | 0.0425 | 473 | 0.0353 |
| 77 | 0.1338 | 177 | 0.0748 | 277 | 0.0535 | 377 | 0.0422 | 477 | 0.0351 |
| 81 | 0.1292 | 181 | 0.0735 | 281 | 0.0529 | 381 | 0.0418 | 481 | 0.0349 |
| 85 | 0.1251 | 185 | 0.0723 | 285 | 0.0523 | 385 | 0.0415 | 485 | 0.0347 |
| 89 | 0.1213 | 189 | 0.0713 | 289 | 0.0518 | 389 | 0.0412 | 489 | 0.0344 |
| 93 | 0.1178 | 193 | 0.0702 | 293 | 0.0513 | 393 | 0.0409 | 493 | 0.0342 |
| 97 | 0.1144 | 197 | 0.0692 | 297 | 0.0507 | 397 | 0.0406 | 497 | 0.0340 |
| 101 | 0.1114 | 201 | 0.0681 | 301 | 0.0501 | 401 | 0.0402 | 501 | 0.0338 |

# Appendix 4

## True optimal weights in scenario $\mathcal{S}_2$

**Table 6.** The true optimal weights for $(q_1+q_3)/2$ in scenario $\mathcal{S}_2$, with $n$ ranging from 5 to 501.

| n | $w_{\text{opt}}(n)$ | n | $w_{\text{opt}}(n)$ | n | $w_{\text{opt}}(n)$ | n | $w_{\text{opt}}(n)$ | n | $w_{\text{opt}}(n)$ |
|---|---|---|---|---|---|---|---|---|---|
| 5 | 0.7786 | 105 | 0.7029 | 205 | 0.7007 | 305 | 0.7004 | 405 | 0.7001 |
| 9 | 0.7436 | 109 | 0.7028 | 209 | 0.7008 | 309 | 0.7001 | 409 | 0.6999 |
| 13 | 0.7301 | 113 | 0.7024 | 213 | 0.7009 | 313 | 0.7004 | 413 | 0.6997 |
| 17 | 0.7225 | 117 | 0.7023 | 217 | 0.7010 | 317 | 0.7004 | 417 | 0.7000 |

**Table 6.** Continued

| n | $w_{opt}(n)$ | n | $w_{opt}(n)$ | n | $w_{opt}(n)$ | n | $w_{opt}(n)$ | n | $w_{opt}(n)$ |
|---|---|---|---|---|---|---|---|---|---|
| 21 | 0.7180 | 121 | 0.7022 | 221 | 0.7009 | 321 | 0.7003 | 421 | 0.6999 |
| 25 | 0.7150 | 125 | 0.7020 | 225 | 0.7006 | 325 | 0.7000 | 425 | 0.6998 |
| 29 | 0.7126 | 129 | 0.7022 | 229 | 0.7007 | 329 | 0.7002 | 429 | 0.6999 |
| 33 | 0.7108 | 133 | 0.7021 | 233 | 0.7006 | 333 | 0.7003 | 433 | 0.6997 |
| 37 | 0.7098 | 137 | 0.7020 | 237 | 0.7005 | 337 | 0.7002 | 437 | 0.7001 |
| 41 | 0.7088 | 141 | 0.7017 | 241 | 0.7008 | 341 | 0.7001 | 441 | 0.6999 |
| 45 | 0.7078 | 145 | 0.7017 | 245 | 0.7006 | 345 | 0.7000 | 445 | 0.7000 |
| 49 | 0.7071 | 149 | 0.7015 | 249 | 0.7006 | 349 | 0.7002 | 449 | 0.7001 |
| 53 | 0.7066 | 153 | 0.7014 | 253 | 0.7007 | 353 | 0.7000 | 453 | 0.6999 |
| 57 | 0.7060 | 157 | 0.7014 | 257 | 0.7004 | 357 | 0.7000 | 457 | 0.6997 |
| 61 | 0.7055 | 161 | 0.7014 | 261 | 0.7008 | 361 | 0.7001 | 461 | 0.6998 |
| 65 | 0.7049 | 165 | 0.7012 | 265 | 0.7006 | 365 | 0.7000 | 465 | 0.6998 |
| 69 | 0.7046 | 169 | 0.7012 | 269 | 0.7004 | 369 | 0.7002 | 469 | 0.7000 |
| 73 | 0.7045 | 173 | 0.7014 | 273 | 0.7003 | 373 | 0.7000 | 473 | 0.6998 |
| 77 | 0.7041 | 177 | 0.7013 | 277 | 0.7003 | 377 | 0.7000 | 477 | 0.6998 |
| 81 | 0.7037 | 181 | 0.7012 | 281 | 0.7004 | 381 | 0.7001 | 481 | 0.6997 |
| 85 | 0.7038 | 185 | 0.7010 | 285 | 0.7004 | 385 | 0.7000 | 485 | 0.6997 |
| 89 | 0.7034 | 189 | 0.7010 | 289 | 0.7005 | 389 | 0.7001 | 489 | 0.6997 |
| 93 | 0.7033 | 193 | 0.7008 | 293 | 0.7003 | 393 | 0.7001 | 493 | 0.6998 |
| 97 | 0.7031 | 197 | 0.7010 | 297 | 0.7002 | 397 | 0.7000 | 497 | 0.6996 |
| 101 | 0.7028 | 201 | 0.7009 | 301 | 0.7003 | 401 | 0.7000 | 501 | 0.6997 |

# Appendix 5

# True optimal weights in scenario $\mathcal{S}_3$

**Table 7.** The true optimal weights $w_{1opt}(n)$ and $w_{2opt}(n)$ in scenario $\mathcal{S}_3$, with n ranging from 5 to 501.

| n | $w_{1opt}(n)$ | $w_{2opt}(n)$ | n | $w_{1opt}(n)$ | $w_{2opt}(n)$ | n | $w_{1opt}(n)$ | $w_{2opt}(n)$ | n | $w_{1opt}(n)$ | $w_{2opt}(n)$ |
|---|---|---|---|---|---|---|---|---|---|---|---|
| 5 | 0.4000 | 0.4000 | 133 | 0.0553 | 0.6556 | 257 | 0.0343 | 0.6722 | 381 | 0.0254 | 0.6795 |
| 9 | 0.2917 | 0.4760 | 137 | 0.0541 | 0.6567 | 261 | 0.0339 | 0.6726 | 385 | 0.0252 | 0.6796 |
| 13 | 0.2386 | 0.5154 | 141 | 0.0532 | 0.6571 | 265 | 0.0334 | 0.6731 | 389 | 0.0251 | 0.6795 |
| 17 | 0.2053 | 0.5403 | 145 | 0.0521 | 0.6585 | 269 | 0.0331 | 0.6733 | 393 | 0.0248 | 0.6797 |
| 21 | 0.1819 | 0.5579 | 149 | 0.0511 | 0.6590 | 273 | 0.0328 | 0.6735 | 397 | 0.0246 | 0.6797 |
| 25 | 0.1643 | 0.5713 | 153 | 0.0501 | 0.6601 | 277 | 0.0324 | 0.6739 | 401 | 0.0244 | 0.6800 |
| 29 | 0.1503 | 0.5822 | 157 | 0.0492 | 0.6606 | 281 | 0.0320 | 0.6743 | 405 | 0.0243 | 0.6804 |
| 33 | 0.1391 | 0.5905 | 161 | 0.0483 | 0.6616 | 285 | 0.0318 | 0.6740 | 409 | 0.0241 | 0.6804 |
| 37 | 0.1298 | 0.5980 | 165 | 0.0475 | 0.6620 | 289 | 0.0313 | 0.6746 | 413 | 0.0239 | 0.6806 |
| 41 | 0.1217 | 0.6042 | 169 | 0.0467 | 0.6625 | 293 | 0.0310 | 0.6750 | 417 | 0.0237 | 0.6808 |
| 45 | 0.1147 | 0.6097 | 173 | 0.0459 | 0.6632 | 297 | 0.0308 | 0.6750 | 421 | 0.0235 | 0.6804 |
| 49 | 0.1087 | 0.6147 | 177 | 0.0451 | 0.6638 | 301 | 0.0305 | 0.6751 | 425 | 0.0234 | 0.6806 |
| 53 | 0.1032 | 0.6185 | 181 | 0.0444 | 0.6646 | 305 | 0.0301 | 0.6753 | 429 | 0.0233 | 0.6809 |
| 57 | 0.0987 | 0.6218 | 185 | 0.0437 | 0.6651 | 309 | 0.0299 | 0.6756 | 433 | 0.0231 | 0.6811 |
| 61 | 0.0944 | 0.6251 | 189 | 0.0430 | 0.6653 | 313 | 0.0296 | 0.6761 | 437 | 0.0229 | 0.6812 |
| 65 | 0.0905 | 0.6286 | 193 | 0.0423 | 0.6664 | 317 | 0.0292 | 0.6763 | 441 | 0.0227 | 0.6812 |
| 69 | 0.0871 | 0.6309 | 197 | 0.0417 | 0.6661 | 321 | 0.0290 | 0.6762 | 445 | 0.0225 | 0.6812 |
| 73 | 0.0836 | 0.6334 | 201 | 0.0411 | 0.6669 | 325 | 0.0287 | 0.6766 | 449 | 0.0224 | 0.6816 |
| 77 | 0.0808 | 0.6360 | 205 | 0.0405 | 0.6675 | 329 | 0.0284 | 0.6771 | 453 | 0.0223 | 0.6820 |
| 81 | 0.0780 | 0.6378 | 209 | 0.0399 | 0.6677 | 333 | 0.0282 | 0.6769 | 457 | 0.0221 | 0.6814 |
| 85 | 0.0755 | 0.6401 | 213 | 0.0394 | 0.6682 | 337 | 0.0279 | 0.6775 | 461 | 0.0219 | 0.6821 |

(continued)

**Table 7.** Continued

| n | $w_{1\text{opt}}(n)$ | $w_{2\text{opt}}(n)$ | n | $w_{1\text{opt}}(n)$ | $w_{2\text{opt}}(n)$ | n | $w_{1\text{opt}}(n)$ | $w_{2\text{opt}}(n)$ | n | $w_{1\text{opt}}(n)$ | $w_{2\text{opt}}(n)$ |
|---|---|---|---|---|---|---|---|---|---|---|---|
| 89 | 0.0732 | 0.6422 | 217 | 0.0389 | 0.6687 | 341 | 0.0277 | 0.6772 | 465 | 0.0218 | 0.6820 |
| 93 | 0.0712 | 0.6433 | 221 | 0.0383 | 0.6691 | 345 | 0.0274 | 0.6778 | 469 | 0.0217 | 0.6821 |
| 97 | 0.0691 | 0.6451 | 225 | 0.0379 | 0.6691 | 349 | 0.0271 | 0.6780 | 473 | 0.0215 | 0.6824 |
| 101 | 0.0671 | 0.6467 | 229 | 0.0374 | 0.6699 | 353 | 0.0269 | 0.6782 | 477 | 0.0214 | 0.6823 |
| 105 | 0.0654 | 0.6479 | 233 | 0.0370 | 0.6700 | 357 | 0.0267 | 0.6782 | 481 | 0.0212 | 0.6823 |
| 109 | 0.0636 | 0.6494 | 237 | 0.0364 | 0.6708 | 361 | 0.0265 | 0.6785 | 485 | 0.0211 | 0.6824 |
| 113 | 0.0621 | 0.6507 | 241 | 0.0360 | 0.6706 | 365 | 0.0263 | 0.6784 | 489 | 0.0210 | 0.6826 |
| 117 | 0.0606 | 0.6516 | 245 | 0.0355 | 0.6714 | 369 | 0.0261 | 0.6788 | 493 | 0.0208 | 0.6828 |
| 121 | 0.0593 | 0.6526 | 249 | 0.0351 | 0.6717 | 373 | 0.0259 | 0.6792 | 497 | 0.0207 | 0.6826 |
| 125 | 0.0578 | 0.6541 | 253 | 0.0346 | 0.6723 | 377 | 0.0256 | 0.6790 | 501 | 0.0206 | 0.6831 |
| 129 | 0.0566 | 0.6548 | | | | | | | | | |

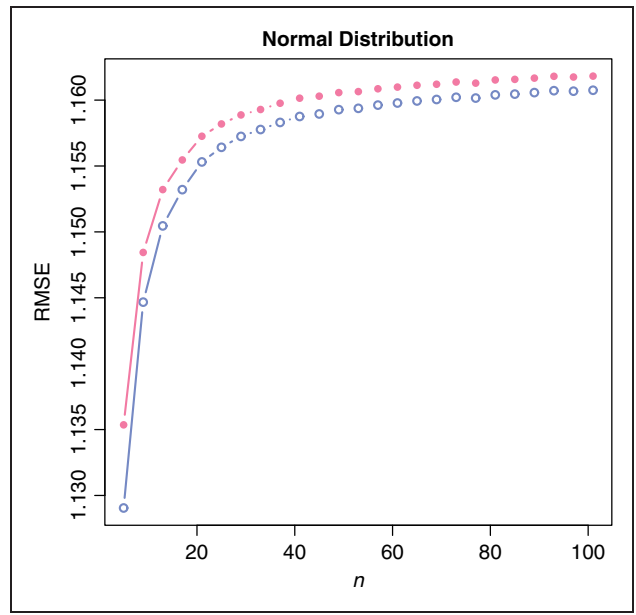

**Figure 4.** RMSE of the sample mean estimation for data from normal distribution for scenario $S_2$. The pink line with solid circles represents Wan et al.'s method, and the light blue line with empty circles represents the new estimator.

# Appendix 6
## Simulation study for $S_2$

In this simulation study, we compare Wan et al.'s estimator (2) and our proposed method (11). For Wan et al.'s method, the three quantities are equally weighted so that the weight to $(q_1 + q_3)/2$ is equal to 0.667. For our new method, we have shown that the optimal weight of $(q_1 + q_3)/2$ should be about 0.7 for moderate to large $n$. Given that the two weights are not far away, we expect that their performance would also be similar. This actually has been demonstrated by the simulation results in Figure 4 for normal data and in Figure 5 for skewed data, based on a total of 100,000,000 simulations. From both figures, we note that our new estimator consistently provides a slightly smaller RMSE than Wan et al.'s estimator, no matter whether $n$ is small or large. We hence conclude that the new estimator is capable to provide a more accurate estimation of the sample mean for both normal and skewed distributions for scenario $S_2$.

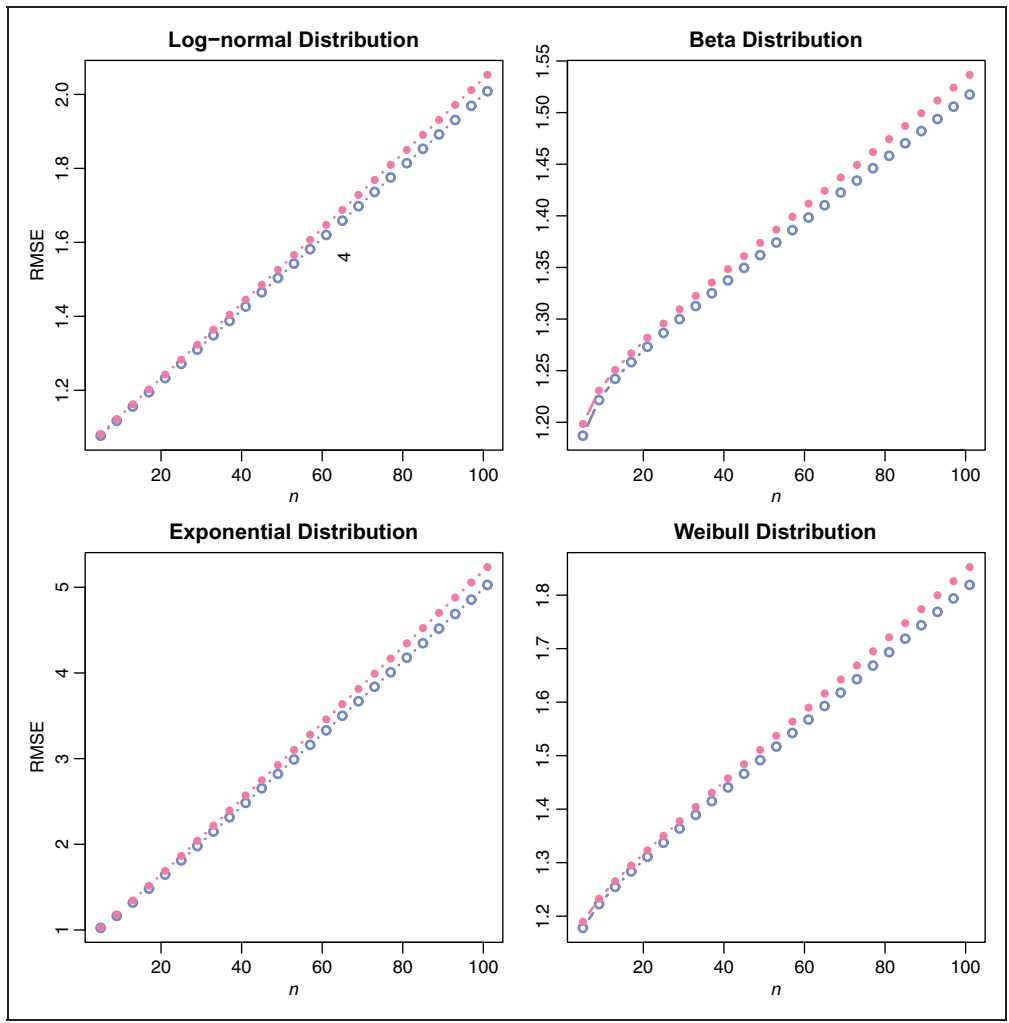

**Figure 5.** RMSE of the sample mean estimation for data from non-normal distributions for scenario $\mathcal{S}_2$. The pink line with solid circles represents Wan et al.'s method, and the light blue line with empty circles represents the new estimator.

# Appendix 7
## Simulation study for $\mathcal{S}_3$

In this simulation study, we compare the performance of Bland's estimator (3) and our proposed estimator (15). Figures 6 and 7 report the results of 100,000 simulations for normal data and skewed data, respectively. It is evident that our new method provides a more stable performance for both normal and skewed data compared with Bland's method. In particular, when $n$ increases, the RMSE of Bland's estimator increases rapidly whereas our estimator provides a relatively stable RMSE that is more close to 1. In view of this, we conclude that our new estimator has provided a more reliable estimate of the sample mean for both normal and skewed distributions for scenario $\mathcal{S}_3$.

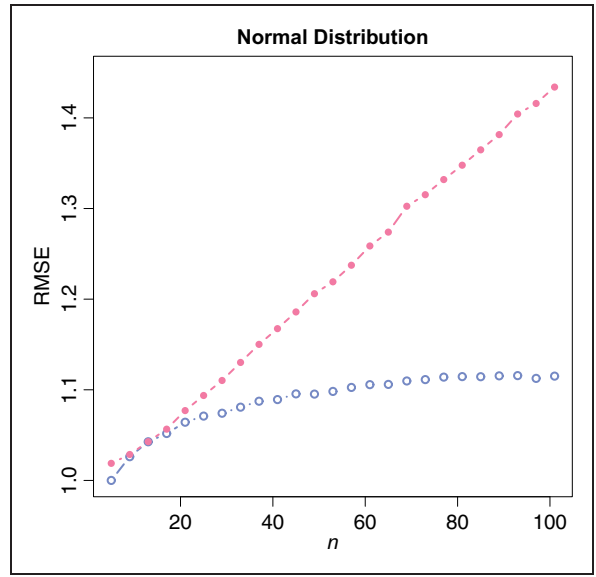

**Figure 6.** RMSE of the sample mean estimation for data from normal distribution for scenario $\mathcal{S}_3$. The pink line with solid circles represents Bland's method, and the light blue line with empty circles represents the new estimator.

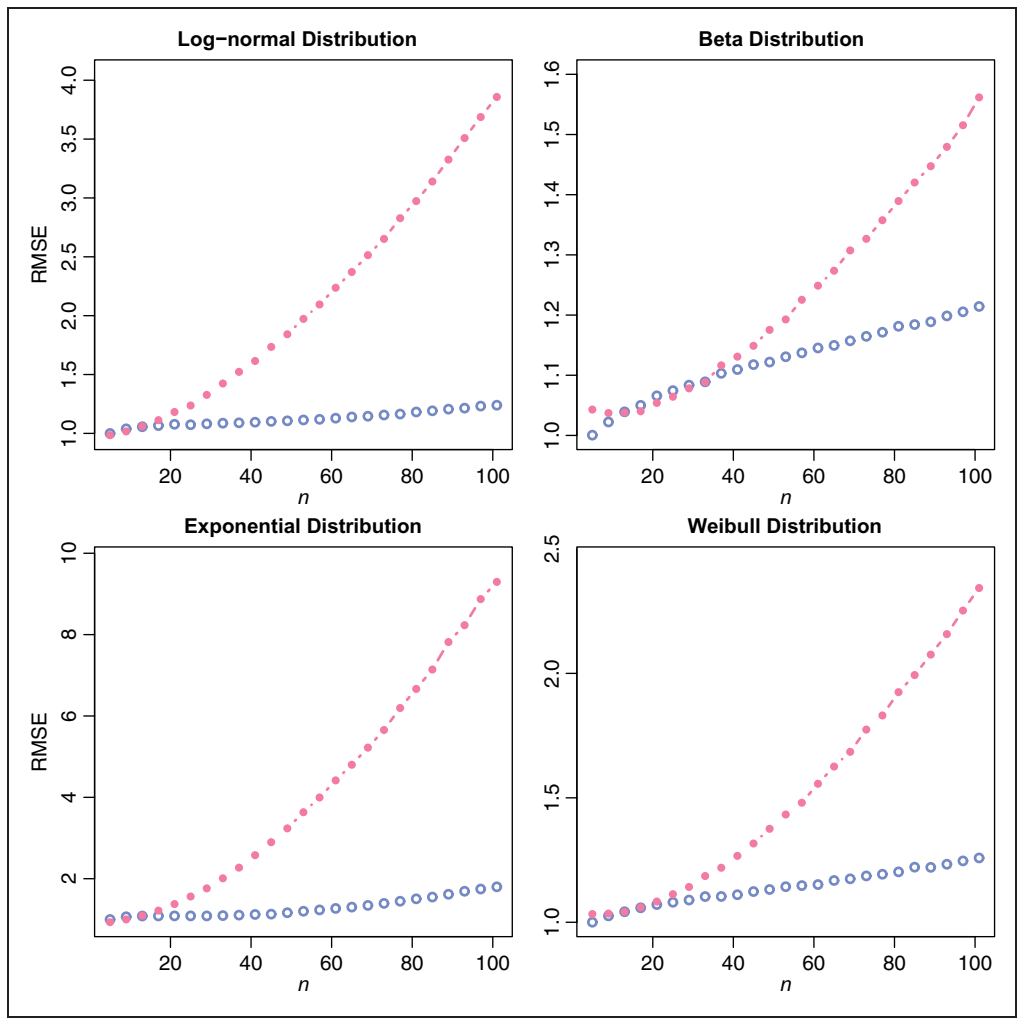

**Figure 7.** RMSE of the sample mean estimation for data from non-normal distributions for scenario $\mathcal{S}_3$. The pink line with solid circles represents Bland's method, and the light blue line with empty circles represents the new estimator.